\documentclass[aps,prb,reprint,floats,citeautoscript,floatfix,nobibnotes,nofootinbib]{revtex4-1}

\usepackage[T1]{fontenc}
\usepackage[english]{babel}
\usepackage{amsmath}
\usepackage{bm}
\usepackage{mathptmx}
\usepackage{braket}

\usepackage{graphicx}

\usepackage{array}
\usepackage{dcolumn}
\newcolumntype{d}[1]{D{.}{.}{#1}}
\usepackage{numprint}

\usepackage{color}
\definecolor{darkred}{rgb}{0.7,0.1,0.1}
\definecolor{darkgreen}{rgb}{0.1,0.6,0.1}
\definecolor{darkblue}{rgb}{0.15,0.15,0.5}
\usepackage[colorlinks=true,allcolors=darkblue,urlcolor=darkblue]{hyperref}

\bibpunct{[}{]}{,}{n}{}{}

\newcommand{\cf}{\mathrm{cf}}
\newcommand{\so}{\mathrm{so}}
\newcommand{\kv}{\mathbf{k}}
\newcommand{\pv}{\mathbf{p}}
\newcommand{\Ev}{\mathbf{E}}
\newcommand{\Gv}{\mathbf{G}}
\newcommand{\eps}{\varepsilon}
\newcommand{\aB}{a_\mathrm{B}}
\newcommand{\kB}{k_\mathrm{B}}
\newcommand{\Angst}{\text{\AA}}

\newcommand{\jena}{Institut f\"ur Festk\"orpertheorie und -optik, Friedrich-Schiller-Universit\"at Jena, Max-Wien-Platz 1, 07743 Jena, Germany and European Theoretical Spectroscopy Facility}

\begin{document}

\title{Accurate electronic and optical properties of hexagonal germanium for optoelectronic applications}

\author{Claudia R\"{o}dl}
\author{J\"{u}rgen Furthm\"{u}ller}
\author{Jens Ren\`{e} Suckert}
\author{Valerio Armuzza}
\author{Friedhelm Bechstedt}
\author{Silvana Botti}
\affiliation{\jena}

\date{\today}

\begin{abstract}
High-quality defect-free lonsdaleite Si and Ge can now be grown on hexagonal nanowire substrates. These hexagonal phases of group-IV semiconductors have been predicted to exhibit improved electronic and optical properties for optoelectronic applications. While lonsdaleite Si is a well-characterized indirect semiconductor, experimental data and reliable calculations on lonsdaleite Ge are scarce and not consistent regarding the nature of its gap. Using \emph{ab initio} density-functional theory, we calculate accurate structural, electronic, and optical properties for hexagonal Ge. Given the well-known sensitivity of electronic-structure calculations for Ge to the underlying approximations, we systematically test the performance of several exchange-correlation functionals, including meta-GGA and hybrid functionals. We first validate our approach for cubic Ge, obtaining atomic geometries and band structures in excellent agreement with available experimental data. Then, the same approach is applied to predict electronic and optical properties of lonsdaleite Ge. We portray lonsdaleite Ge as a direct semiconductor with only weakly dipole-active lowest optical transitions, small band gap, huge crystal-field splitting, and strongly anisotropic effective masses. The unexpectedly small direct gap and the oscillator strengths of the lowest optical transitions are explained in terms of symmetry and back-folding of energy bands of the diamond structure.
\end{abstract}

\maketitle

\section{Introduction}
\label{sec:intro}

The integration of a material featuring efficient interaction with light into Si technology is of high technological interest. In fact, the copper interconnect between transistors on a chip has become a bigger challenge than reducing transistor size. A possible solution to this critical bottleneck are optical interconnects \cite{Miller:2009:PI}. Si in the diamond structure is an indirect band-gap material and cannot be used for this purpose. Several attempts to obtain light emission from Si in the telecommunication band had only limited success \cite{Pavesi.DalNegro.ea:2000:N,Han.Seo.ea:2001:APL,Rong.Liu.ea:2005:N}.

Si and Ge, despite their chemical similarity, are very different from an optical perspective. Ge, one of the most important and widely used semiconductors, crystallizes like Si in the cubic diamond structure (space group $Fd\bar{3}m$) with an indirect band gap of about 0.7~eV under ambient conditions~\cite{Yu.Cardona:1996:Book}. Diamond-structure Ge is characterized by a poor light-emission efficiency because of the indirect nature of its fundamental band gap. However, its direct band-gap energy is close to the indirect one, and significant engineering efforts are being made to convert Ge into an efficient gain material monolithically integrated on a Si chip \cite{Xu.Narusawa.ea:2012:IJoSTiQE,Sun.Liu.ea:2009:APL,Kersauson.Kurdi.ea:2011:OE,Apetz.Vescan.ea:1995:APL}. To raise the interest in Ge for possible active optoelectronic applications, e.g.\ in light-emitting diodes or lasers, the $\kv$-selection rule that forbids optical dipole transitions at the minimum band gap of Ge has to be broken. To this end, ingenious approaches have been proposed, for instance, based on straining~\cite{Zhang.Crespi.ea:2009:PRL}, nanostructuring~\cite{Lockwood.Rowell.ea:2013:ET,Takeoka.Fujii.ea:1998:PRB}, or amorphization~\cite{Laubscher.Kuefner.ea:2015:JoPCM}. 

\begin{figure}
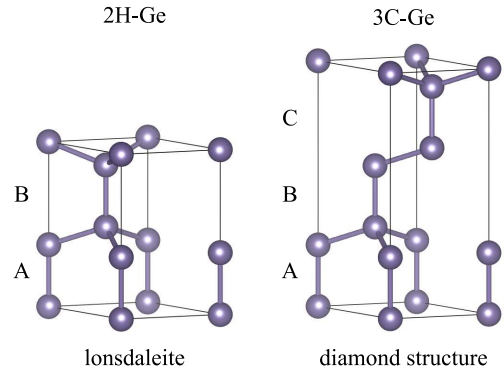

\centering\includegraphics[width=0.8\linewidth]{{{fig1}}}
\caption{\label{fig:structures} Atomic bilayer stacking in the lonsdaleite (2H) and diamond-structure (3C) phase of Ge. The figures of the crystal structures have been produced with \textsc{Vesta}~\cite{Momma.Izumi:2011:JoAC}.}
\end{figure}

Besides the thermodynamically stable diamond structure, other metastable allotropes of Ge have been explored for optical applications. Ge in the hexagonal lonsdaleite structure (space group $P6_3/mmc$), occasionally also called wurtzite Ge, is attracting increasing attention as a promising material for optoelectronics. In the lonsdaleite phase, the Ge atoms feature the same tetrahedral nearest-neighbor coordination as in the cubic diamond structure, but, instead of an ABC stacking of adjacent Ge bilayers along the threefold symmetry axis, the lonsdaleite phase is characterized by an AB stacking~\cite{Raffy.Furthmueller.ea:2002:PRB}. Therefore, we refer to cubic Ge in the diamond structure also as 3C-Ge and to hexagonal Ge in the lonsdaleite structure as 2H-Ge (see Fig.~\ref{fig:structures}). Lonsdaleite Ge was first obtained at low pressure using ultraviolet laser ablation~\cite{Zhang.Iqbal.ea:2000:SSC,Haberl.Guthrie.ea:2014:PRB}. Recently, it has been grown on top of a template of wurtzite-GaP nanowires in form of core-shell nanowires~\cite{Haverkort::tbp}, similarly to what had already been achieved for lonsdaleite Si~\cite{Hauge.Verheijen.ea:2015:NL}. Other routes towards hexagonal Ge nanowires have also been suggested \cite{Jeon.Dayeh.ea:2013:NL,Vincent.Patriarche.ea:2014:NL}. Moreover, exploiting strain-induced phase transformations, Ge nanowires featuring homojunctions of different polytypes can be synthesized~\cite{Vincent.Patriarche.ea:2014:NL}.

In the hexagonal Brillouin zone (BZ) of the lonsdaleite structure, the $L$ point of the diamond-structure BZ that lies on the cubic [111] axis is mapped onto the $\Gamma$~point. Therefore, the lowest conduction-band minimum (CBM) at the $L$~point of cubic Ge is folded onto the $\Gamma$~point rendering 2H-Ge a direct-gap semiconductor. Comparing hexagonal Si and Ge, it seems that breaking the $\kv$-selection rule is easier in Ge, since the original and the backfolded conduction band are energetically very close ~\cite{Weissker.Furthmuller.ea:2002:PRB}. The exact ordering of the lowest conduction bands at $\Gamma$ is extremely important, as the electron radiative lifetime of the material strongly depends on the symmetry of these states. 

Together with optical emission or absorption measurements for photon energies comparable with the size of the band gap, accurate electronic-structure calculations can provide detailed answers concerning conduction-band ordering and the strength of optical transitions. Despite the fact that Ge is an elemental material, the experience acquired with calculations of diamond-structure Ge proves that Ge is a difficult system for accurate band-structure studies~\cite{Cartoixa.Palummo.ea:2017:NL,Kaewmaraya.Vincent.ea:2017:JPCC,Hybertsen.Louie:1986:PRB,Chen.Fan.ea:2017:JoPDAP}. On the one hand, it is essential to account for spin-orbit coupling (SOC) and to treat the shallow Ge $3d$ shell as valence electrons~\cite{Laubscher.Kuefner.ea:2015:JoPCM}. On the other hand, the approximations used to describe exchange and correlation (XC) contributions to the electron-electron interaction significantly influence the $\kv$-space position of the lowest CBM and the size of the direct and indirect gaps (see for example the discussion in Ref.~\cite{Laubscher.Kuefner.ea:2015:JoPCM}). 

Applying density-functional theory (DFT)~\cite{Hohenberg.Kohn:1964,Kohn.Sham:1965}, a Kohn-Sham (KS) band structure obtained within the local-density approximation (LDA) or any flavor of the generalized gradient approximation (GGA) is not sufficient, as the system is erroneously predicted to be metallic in this case~\cite{Raffy.Furthmueller.ea:2002:PRB,Laubscher.Kuefner.ea:2015:JoPCM,Wang.Ye:2003:JoPCM,Cartoixa.Palummo.ea:2017:NL,Waroquiers.Lherbier.ea:2013:PRB}. Moreover, the band gap is extremely sensitive to the value of the lattice constant. Therefore, an accurate description of the atomic geometry is indispensable. More sophisticated approaches, beyond semi-local XC functionals, are needed to obtain reliable quasiparticle states~\cite{Bechstedt:2015:Book}. We will focus here on the question whether poor-men's approaches using hybrid DFT functionals~\cite{Laubscher.Kuefner.ea:2015:JoPCM,Kaewmaraya.Vincent.ea:2017:JPCC} or meta-GGA functionals~\cite{Laubscher.Kuefner.ea:2015:JoPCM,Waroquiers.Lherbier.ea:2013:PRB} can be sufficient. 

In fact, 3C-Ge band structures have also been computed approximating the XC self-energy within Hedin's $GW$ approximation~\cite{Hybertsen.Louie:1986:PRB,Waroquiers.Lherbier.ea:2013:PRB,Chen.Fan.ea:2017:JoPDAP}, and we can expect this state-of-the-art approach for excited states to work also for lonsdaleite Ge. However, the computational cost coming along with Green's function calculations is very high and would likely make their application to more complex systems (such as alloys, doped or defective crystals, surfaces, or interfaces) unfeasible.  We remark that also the empirical-pseudopotential method (EPM), widely used for 3C-Ge, can be helpful at a reduced computational cost~\cite{Joannopoulos.Cohen:1973:PRB,De.Pryor:2014:JoPCM}.  However, empirical approaches that very accurately reproduce experimental data for 3C-Ge would need additional assumptions to be reliably applied to 2H-Ge.

In this paper, we present a careful analysis of the electronic and optical properties of 2H-Ge, with a particular focus on the choice of accurate and computationally efficient XC functionals for ground-state and excited-state calculations. The functionals are first tested against experimental data for 3C-Ge. They are then used for a careful analysis of the electronic and optical properties of 2H-Ge. In view of potential optoelectronic applications, we are especially interested in conduction-band ordering, direct and indirect band gaps, band splittings, effective masses, optical transition strengths, and radiative lifetimes. The methods used are described in Sec.~\ref{sec:methods}. In Sec.~\ref{sec:cub}, we discuss the applicability and efficiency of different approximations for 3C-Ge. The results are then used in Sec.~\ref{sec:hex} to carefully analyze the electronic and optical properties of 2H-Ge. Finally, in Sec.~\ref{sec:summary}, we present a summary of our results and draw conclusions.


\section{Methods}
\label{sec:methods}

\subsection{Ground-state calculations}

All calculations were performed with the Vienna Ab-initio Simulation Package (\textsc{Vasp})~\cite{Kresse.Furthmuller:1996} with the projector-augmented wave (PAW) method~\cite{Kresse.Joubert:1999:PRB} and a plane-wave cutoff of 500~eV. The shallow Ge~$3d$ electrons were explicitly included as valence electrons. BZ integrations were carried out using $12\times12\times12$ (3C-Ge) or $12\times12\times6$ (2H-Ge) $\Gamma$-centered $\kv$~points (unless otherwise stated), ensuring a convergence of total energies to 1~meV/atom. Atomic geometries and elastic properties were calculated with (semi-)local XC functionals, using the LDA~\cite{Kohn.Sham:1965} as well as the GGA flavors PBE~\cite{Perdew.Burke.ea:1996}, PBEsol~\cite{Perdew.Ruzsinszky.ea:2008:PRL} (a modified version of the PBE functional optimized for solids), and AM05~\cite{Armiento.Mattsson:2005:PRB,Mattsson.Armiento.ea:2008:TJoCP}.

The ground-state atomic structures, the isothermal bulk modulus $B_0$, and its pressure derivative $B_0'$ were determined by a series of fixed-volume relaxations and a subsequent fit of the resulting energy-over-volume curve to the Vinet equation of state (EOS)~\cite{Vinet.Ferrante.ea:1986:JoPCSSP,Vinet.Rose.ea:1989:JoPCM}. The internal cell parameters were relaxed until the Hellmann-Feynman forces drop below 1~meV/\AA. We found that the inclusion of SOC has essentially no impact on the lattice parameters and only a minor effect on the elastic constants. This observation is in line with general conclusions for other simple solids and zincblende-type semiconductors~\cite{DalCorso:2012:PRB}.


\subsection{Electronic structures}

It is well known that KS band structures calculated in the LDA or GGA significantly underestimate all band gaps and interband transition energies~\cite{Guilloy.Pauc.ea:2015:NL,Bechstedt:2015:Book}. Quasiparticle calculations in the state-of-the-art $GW$ approximation, on the other hand, are challenging and computer-time consuming for Ge, due to the necessity to include SOC, to account for the $3d$ electrons,  and to calculate the full dynamical screening.

What is more, the $GW$ quasiparticle band structures need to be computed self-consistently to overcome the problem of the negative fundamental gap in the LDA/GGA starting electronic structure of both 3C-Ge and 2H-Ge (see Sections~\ref{sec:cub} and \ref{sec:hex}). One reason for the negative gaps is the overestimation of the $p$-$d$ repulsion~\cite{Wei.Zunger:1988:PRB}. This is a direct consequence of the underestimated  binding energy of the Ge~$3d$ electrons within LDA or GGA, which pushes the $p$-like valence-band maximum (VBM) towards higher energies. An improved description of the localized $d$ states can be achieved within the DFT+$U$ method~\cite{Anisimov.Zaanen.ea:1991} with a Hubbard parameter $U$ for the $3d$ electrons.  We tested the DFT+$U$ method in the Dudarev approach~\cite{Dudarev.Botton.ea:1998:PRB} using a small but reasonable value $U=1.3$~eV, which is in rough agreement with the picture of an atomic Coulomb integral of about $U^\mathrm{atom}=15$~eV screened by the bulk Ge dielectric constant \cite{Harrison:1999:Book}.

We further used the HSE06 hybrid functional ~\cite{Heyd.Scuseria.ea:2003,Krukau.Vydrov.ea:2006:JCP} with a fraction $\alpha=0.25$ of short-range Fock exchange and an inverse screening length $\omega=0.2~\Angst^{-1}$ to calculate reliable band structures for cubic and hexagonal Ge. It has been shown that the HSE06 functional yields reasonable indirect and direct gaps for Ge~\cite{Laubscher.Kuefner.ea:2015:JoPCM,Kaewmaraya.Vincent.ea:2017:JPCC} and many other $sp$ semiconductors~\cite{Kaewmaraya.Vincent.ea:2017:JPCC,Fuchs.Furthmuller.ea:2007:PRB,Kim.Hummer.ea:2009:PRB,Marques.Vidal.ea:2011:PRB,Bechstedt:2015:Book}. The most important contribution to the gap opening within the $GW$ approach is due to the screened-exchange part of the electronic self-energy. The Coulomb hole, the second contribution to the $GW$ self-energy, mainly influences the absolute position of the one-particle energies \cite{Bechstedt:2015:Book}. In the HSE06 functional, the fraction $\alpha$ of Fock exchange simulates the important non-locality feature of the self-energy and the screening of the electron-electron interaction by an average dielectric constant of $1/\alpha$ ~\cite{Bechstedt:2015:Book,Marques.Vidal.ea:2011:PRB}.

As a computationally cheap alternative to hybrid functionals, we also consider the meta-GGA functional~\cite{Perdew.Ruzsinszky.ea:2005:JCP} MBJLDA of Tran and Blaha~\cite{Tran.Blaha.ea:2007:JoPCM,Tran.Blaha:2009:PRL} that is based on the modified Becke-Johnson (MBJ) exchange functional~\cite{Becke.Johnson:2006:JCP}. The MBJLDA functional does not only give reasonable band gaps for 3C-Ge \cite{Laubscher.Kuefner.ea:2015:JoPCM} but also for other semiconductors~\cite{Kim.Marsman.ea:2010:PRB,Waroquiers.Lherbier.ea:2013:PRB}. The strongly reduced computational cost allows for the application of the MBJLDA functional to more complex systems. In particular in the context of potential optoelectronic applications of 2H-Ge, also strained, disordered, or defective systems with larger supercells become computationally accessible. Moreover, both the hybrid and the meta-GGA functional allow for an easy inclusion of SOC.


\subsection{Optical properties}

Having in mind optoelectronic applications (e.g.\ lasing), the global optical emission properties of 2H-Ge near the fundamental absorption edge can be characterized by the optical transition matrix elements of the near-edge transitions and the radiative lifetime of the material. Here, the optical transition matrix elements are calculated in the longitudinal gauge~\cite{Gajdos.Hummer.ea:2006:PRB}. They are given as matrix elements $\braket{c\kv|\pv|v\kv}$ of the momentum operator $\pv$ between conduction band $c$ and valence band $v$ at a given $\kv$~point. 

The optical matrix elements at the $\Gamma$~point can be linked to characteristic quantities from $\kv\cdot\pv$ perturbation theory~\cite{Yu.Cardona:1996:Book} introducing the average of the squared momentum matrix element over spin-orbit degenerate states $i,j=1,2$ in the conduction and valence bands at the zone center,
\begin{equation}\label{eq:optmat}
p^{\perp/\|} = \sqrt{\frac{1}{2} \sum_{c_i,v_j} \left|\braket{c_i\kv=0|p^{\perp/\|}|v_j\kv=0}\right|^2}.
\end{equation}
Then, the Kane energy reads
\begin{equation}\label{eq:kane}
 E_p^{\perp/\parallel} = \frac{2}{m} \left(p^{\perp/\|}\right)^2
\end{equation}
and the (dimensionless) optical oscillator strength
\begin{equation}\label{eq:oscistrength}
f^{\perp/\|}=\frac{E_p^{\perp/\|}}{\epsilon_{c\kv=0}-\epsilon_{v\kv=0}},
\end{equation}
where $\perp/\|$ stands for light polarized perpendicular/parallel to the $c$~axis of the lonsdaleite structure. For 3C-Ge, these two directions are obviously equivalent.

The radiative lifetime $\tau$ at temperature $T$, as a global measure for the light-emission properties of a material, is given by the thermally averaged recombination rate~\cite{Dexter:1958,Delerue.Allan.ea:1993:PRB} 
\begin{equation}\label{eq:av_rec_rate}
\frac{1}{\tau}= \sum\limits_{cv\kv} A_{cv\kv}\,
\frac{\,w_\kv\, e^{-(\epsilon_{c\kv}-\epsilon_{v\kv})/(\kB T)}}
{\sum\limits_{cv\kv} w_\kv\, e^{-\left(\epsilon_{c\kv}-\epsilon_{v\kv}\right)/(\kB T)}},
\end{equation}
where $A_{cv\kv}$ denotes the radiative recombination rate for vertical optical transitions between a conduction state $\ket{c\kv}$ and a valence state $\ket{v\kv}$ with the one-particle energies $\epsilon_{c\kv}$ and $\epsilon_{v\kv}$ and $\kv$-point weight $w_\kv$. The radiative recombination rate reads
\begin{equation} \label{eq:Acvk}
\begin{split}
A_{cv\kv} = & n_\mathrm{eff}\, \frac{e^2\, (\epsilon_{c\kv}-\epsilon_{v\kv})}{\pi \epsilon_0\,\hbar^2 m^2 c^3} 
\, \frac{1}{3}\sum\limits_{j=x, y, z} \left|\braket{c\kv|p_j|v\kv}\right|^2,
\end{split}
\end{equation}
with $n_\mathrm{eff}$ the refractive index of the effective medium consisting of the Ge sample and its environment (set to 1 in the following). The squares of the momentum matrix elements are averaged over all directions corresponding to the emission of unpolarized light. We stress two important points. First, Eq.~\eqref{eq:Acvk} is given in the independent-(quasi)particle approximation~\cite{Adolph.Gavrilenko.ea:1996}, i.e., neglecting excitonic effects, which, however, can be easily taken into account \cite{Palummo.Bernardi.ea:2015:NL}. Second, in Eq.~\eqref{eq:av_rec_rate}, it is assumed that the thermalization of electrons and holes after their injection is more efficient than the radiative (or non-radiative) recombination~\cite{Delerue.Allan.ea:1993:PRB,Ramos.Furthmuller.ea:2005:PRBa,Allan.Delerue.ea:2001:PRB}. Whereas the convergence of the radiative lifetimes with the number of bands is very fast, we need $72\times72\times72$ (3C-Ge) or $72\times72\times36$ (2H-Ge) $\kv$~points to sample the BZ with sufficiently high accuracy.


\section{Diamond-structure germanium: Validation of the approach}
\label{sec:cub}

\begin{table}[t]
\caption{\label{tab:cub_ground} Lattice constant $a_0$, isothermal bulk modulus $B_0$, its pressure derivative $B_0'$, and cohesive energy $E_\mathrm{coh}$ of 3C-Ge. Experimental (Exp.) values are given for comparison.}
\begin{ruledtabular}
\begin{tabular}{l*{4}{c}}
Method & $a_0$ (\AA) & $B_0$ (GPa) & $B_0'$ & $E_\mathrm{coh}$ (eV/at.) \\ \hline
LDA        & 5.626 & 71.8 & 4.92 & 4.63 \\
PBE        & 5.760 & 58.7 & 5.01 & 3.73 \\
PBEsol     & 5.673 & 67.3 & 4.89 & 4.15 \\
AM05       & 5.677 & 65.7 & 4.82 & 3.92 \\
PBEsol+$U$ & 5.652 & 68.7 & 4.86 & 4.15 \\
AM05+$U$   & 5.656 & 67.2 & 4.87 & 3.91 \\ \hline
Exp.       & 5.658\footnote{X-ray diffraction at $T=298.15~\text{K}$ \cite{Smakula.Kalnajs:1955:PR}.}     
           & 75.0\footnote{Obtained from ultrasonic measurements of the elastic moduli $C_{11}$ and $C_{12}$ at ambient pressure at $T=298.15~\text{K}$ \cite{McSkimin.Andreatch:1963:JoAP} using the relation $B_0=(C_{11}+2C_{12})/3$.}  
           &             
           & 3.85\footnote{From Ref.~\cite{Kittel:2005:Book}.} \\ 
           & 5.652\footnote{X-ray diffraction at $T=10~\text{K}$ \cite{Hu.Sinn.ea:2003:PRB}.}
           & 64.7\footnote{From fitting an EOS to room-temperature experimental data for various pressures \cite{Queisser.Holzapfel:1991:APA}.}
           & 5.0(1)\footnotemark[5]
           & \\
           &
           & 77(4)\footnote{From fitting the Vinet EOS to room-temperature experimental data for various pressures \cite{DiCicco.Frasini.ea:2003:pssb}.}
           & 4.3(1.0)\footnotemark[6]
           &  \\
\end{tabular}
\end{ruledtabular}
\end{table}

The lattice constant, elastic properties, and cohesive energy of 3C-Ge have been calculated with various XC correlation functionals (see Table~\ref{tab:cub_ground}). Comparing with experimental values~\cite{Smakula.Kalnajs:1955:PR,McSkimin.Andreatch:1963:JoAP,Kittel:2005:Book,Queisser.Holzapfel:1991:APA,DiCicco.Frasini.ea:2003:pssb,Hu.Sinn.ea:2003:PRB}, the expected tendencies are visible~\cite{Bechstedt:2015:Book}: The LDA tends to overbind, whereas the inclusion of gradient corrections, in particular within PBE, leads to an underestimation of the strength of the chemical bonds. The functionals PBEsol and AM05 yield the best agreement with experiment. However, they still slightly overestimate the experimental lattice constant. Further improvement can be obtained by DFT+$U$ calculations (see Table~\ref{tab:cub_ground}) at the price of introducing the adjustable parameter $U$. Also the isothermal bulk modulus $B_0$, its pressure derivative $B_0'$, and the cohesive energy are consistent with experiment.

\begin{figure*}
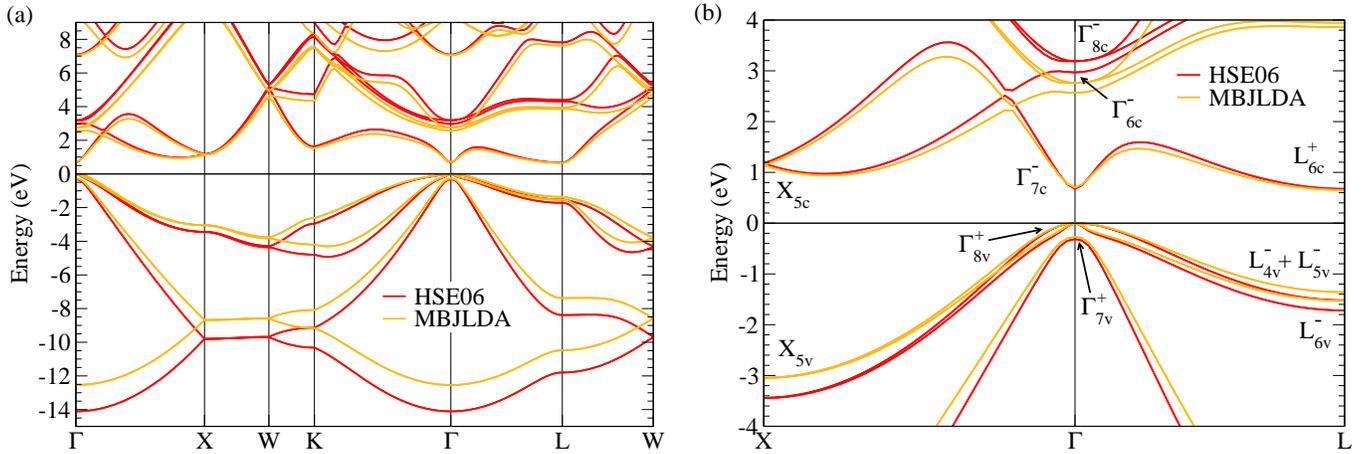

\includegraphics[width=0.49\linewidth]{{fig2a}}
\hfill
\includegraphics[width=0.49\linewidth]{{fig2b}}
\caption{\label{fig:cub_bst} Band structure of 3C-Ge computed with the HSE06 and MBJLDA functionals (a). The irreducible representations of relevant high-symmetry states in the band-gap region are given in the double-group notation of Koster \emph{et al.}~\cite{Koster.Dimmock.ea:1963:Book} (b). The VBM is set to zero.}
\end{figure*}

\begin{table*}[t]
\caption{\label{tab:cub_eigenvalues} Band energies and spin-orbit induced band splittings $\Delta_\so$ of 3C-Ge at high-symmetry points of the BZ calculated with the PBEsol, HSE06, and MBJLDA functionals at the PBEsol lattice constant. Experimental low- and room-temperature values are provided for comparison. All values in eV.}  
\begin{ruledtabular}
\begin{tabular}{l*{5}{c}}
State  & PBEsol & HSE06 & MBJLDA  &  \multicolumn{2}{c}{Experiment} \\ 
 & & & & low temperature & room temperature \\ \hline
$\Gamma_{7v}^+$ & -0.292 & -0.314 & -0.270 
& -0.297\footnote{Schottky-barrier electroreflectance at 10~K \cite{Aspnes:1975:PRB}.} & \\
$\Gamma_{8v}^+$ &  0.000 &  0.000 &  0.000 & & \\
$\Gamma_{7c}^-$ & -0.174 &  0.678 &  0.705 
& 0.887(1)\footnotemark[1], 
  0.898(1)\footnote{Magnetoabsorption at 1.5~K and 293~K~\cite{Zwerdling.Lax.ea:1959:PR}.} 
& 0.805(1)\footnotemark[2] \\
$\Gamma_{6c}^-$ &  2.296 &  2.969 &  2.564 & & \\
$\Gamma_{8c}^-$ &  2.508 &  3.189 &  2.760  
& $\Gamma^-_{6c}+0.200$\footnotemark[1] & \\
$\Delta_\so(\Gamma_{8v}^+-\Gamma_{7v}^+)$ & 0.292 & 0.341 & 0.270 
& 0.297\footnotemark[1] & \\
$\Delta_\so(\Gamma_{8c}^--\Gamma_{6c}^-)$ & 0.212 & 0.220 & 0.196  
& 0.200\footnotemark[1]& \\ \hline
$L_{6v}^-$         & -1.585 & -1.719 & -1.530 & & \\
$L_{4v}^-+L_{5v}^-$ & -1.401 & -1.519 & -1.359 
& $L^-_{6v}+0.228$\footnotemark[1] & \\
$L_{6c}^+$      &  0.005 &  0.675 &  0.653 
& 0.744(1)\footnotemark[2]          
& 0.6643\footnote{Optical absorption-edge fine structure at 291~K \cite{Macfarlane.McLean.ea:1957:PR}.} \\
$\Delta_\so(L_{4v+5v}^--L_{6v}^-)$ & 0.184 & 0.200 & 0.171 
& 0.228\footnotemark[1] & \\ \hline
$X_{5v}$        & -3.155 & -3.441 & -3.047 & & \\
$X_{5c}$        &  0.585 &  1.177 &  1.142 & & \\
\end{tabular}
\end{ruledtabular}
\end{table*}

Subsequently, the band structure of 3C-Ge including SOC was calculated using the PBEsol, HSE06, and MBJLDA functionals at the PBEsol lattice constant (see Fig.~\ref{fig:cub_bst} and Table~\ref{tab:cub_eigenvalues}). The states at the high-symmetry points of the BZ are labeled according to the double-group notation of Koster \emph{et al.}~\cite{Koster.Dimmock.ea:1963:Book}. The small difference between the PBEsol and the experimental lattice constant corresponds to an isotropic tensile strain of $<0.4\,\%$. The volume deformation potential for the direct gap (the most volume-sensitive band-to-band transition) amounts to $-9.0$~eV (MBJLDA) implying that differences in the direct gap due to the discrepancy in the lattice constant are smaller than 0.1~eV. For the sake of comparability, all calculations of electronic and optical properties presented in the following are based on the PBEsol lattice constant. 

GGA functionals like PBEsol yield a negative KS band gap for 3C-Ge in contradiction to experimental findings (cf.\ Table~\ref{tab:cub_eigenvalues}) which is why they are unsuitable for the description of the electronic structure of this material. The band ordering, band energies, and spin-orbit splittings $\Delta_\so$ obtained with the more sophisticated HSE06 functional agree well with experimental results. Comparing HSE06 and MBJLDA band structures close to the fundamental gap, we find similar indirect ($\Gamma_{8v}^+\to L_{6c}^+$) and direct ($\Gamma_{8v}^+\to\Gamma_{7c}^-$) band gaps. Also the spin-orbit splittings of the $p$ states are much the same. Further away from the band-gap region, the discrepancy between HSE06 and MBJLDA band energies increases. This is, however, not a crucial problem here, since we are mostly interested in optoelectronic properties that are governed by the electronic structure in the vicinity of the band gap. In particular, the ordering of the $\Gamma_{7c}^-$ and $L_{6c}^+$ conduction-band minima is correct, only their energy distance is slightly underestimated compared to experiment (independent of temperature). Note that $GW$ corrections on top of HSE06 or MBJLDA band structures are known to overestimate the gaps~\cite{Waroquiers.Lherbier.ea:2013:PRB}. 

\begin{table}[t]
\caption{\label{tab:cub_effmass} Effective electron and hole masses of 3C-Ge in units of the free electron mass $m$. The VBM at $\Gamma$ splits into a heavy hole ($m_\mathrm{h}^\mathrm{hh}$), light hole ($m_\mathrm{h}^\mathrm{lh}$), and spin-orbit split-off hole ($m_\mathrm{h}^\so$). The heavy-hole and light-hole masses are averaged over the $\Gamma$-$X$ and $\Gamma$-$L$ directions. The masses of the CBM at $L$ are given both parallel ($m_\mathrm{e}^\|$) and perpendicular ($m_\mathrm{e}^\perp$) to the $L$-$\Gamma$ direction.}
\begin{ruledtabular}
\begin{tabular}{lccc}
Mass & HSE06 & MBJLDA  & Exp. \\ \hline
$m_\mathrm{h}^\so(\Gamma_{7v}^+)$ & 0.097 & 0.122 
& 0.095(7)\footnote{Piezomagnetoreflectance at 30~K \cite{Aggarwal:1970:PRB}.} \\ 
$m_\mathrm{h}^\mathrm{lh}(\Gamma_{8v}^+)$ & 0.043 & 0.059 
& 0.0438(30) $\mathbf{B}\|[100]$\footnote{Cyclotron resonance at 4~K with magnetic field $\mathbf{B}$ oriented in various directions \cite{Dexter.Zeiger.ea:1956:PR}.} \\
&&& 0.0426(20) $\mathbf{B}\|[111]$\footnotemark[2] \\
&&& 0.0430(30) $\mathbf{B}\|[110]$\footnotemark[2] \\
$m_\mathrm{h}^\mathrm{hh}(\Gamma_{8v}^+)$ & 0.203 & 0.233 & 0.284(1) $\mathbf{B}\|[100]$\footnotemark[2] \\
&&& 0.376(1) $\mathbf{B}\|[111]$\footnotemark[2] \\
&&& 0.352(4) $\mathbf{B}\|[110]$\footnotemark[2] \\
$m_\mathrm{e}(\Gamma_{7c}^-)$ & 0.034 & 0.047 
& 0.0380(5)\footnotemark[1] \\ \hline
$m_\mathrm{e}^\|(L_{6c}^+)$   & 1.573 & 1.728 
& 1.588(5)\footnote{Cyclotron resonance at 1.4~K \cite{Levinger.Frankl:1961:JPCS}.},
  1.59\footnote{Magnetophonon resonance at 120~K \cite{Hirose.Shimomae.ea:1982:JPSJ}.} \\
$m_\mathrm{e}^\perp(L_{6c}^+)$ & 0.090 & 0.096 
& 0.08152(8)\footnotemark[3],
  0.0823\footnotemark[4] \\
\end{tabular}
\end{ruledtabular}
\end{table}

In Table~\ref{tab:cub_effmass}, the electron and hole effective masses of relevant band extrema are compiled. Besides the band masses at the $\Gamma$ point, also the masses of the CBM at the $L$ point parallel and perpendicular to the $L$-$\Gamma$ line are given. The masses have been derived from the corresponding HSE06 and MBJLDA band structures. The HSE06 masses are in excellent agreement with experimental values~\cite{Aggarwal:1970:PRB,Dexter.Zeiger.ea:1956:PR,Levinger.Frankl:1961:JPCS,Hirose.Shimomae.ea:1982:JPSJ}. The MBJLDA functional slightly overestimates the experimental band masses which is in line with previous observations~\cite{Kim.Marsman.ea:2010:PRB} and the generally lower band widths in the MBJLDA calculation compared to the HSE06 calculation (cf.\ Fig.~\ref{fig:cub_bst}).

Considering the findings for 3C-Ge, we rely on the PBEsol functional for the structural properties of 2H-Ge. The HSE06 and MBJLDA functionals will be used to study the electronic and optical properties of lonsdaleite Ge. This strategy is corroborated by the fact that both allotropes of Ge feature similar chemical bonding properties, i.e.\ they are both insulators with tetrahedral coordination. Therefore, the performance of the functionals should be largely transferable.


\section{Lonsdaleite germanium: Predictions}
\label{sec:hex}

\subsection{Atomic geometry and bonding}

\begin{table*}[t]
\caption{\label{tab:hex_ground} Structural and elastic properties of 2H-Ge. Structural parameters $a$, $c$, and $u$ as wells as the isothermal bulk modulus $B_0$, its pressure derivative $B_0'$, and the cohesive energy $E_\mathrm{coh}$. Available experimental data are given for comparison.}
\begin{ruledtabular}
\begin{tabular}{l*{7}{c}}
Method     & $a$ (\AA) & $c$ (\AA) & $c/a$ & $u$ & $B_0$ (GPa) & $B_0'$ & $E_\mathrm{coh}$ (eV/at.) \\
    \hline
LDA        & 3.962 & 6.539 & 1.6504 & 0.3742 & 71.9 & 5.00 & 4.61 \\
PBE        & 4.058 & 6.692 & 1.6492 & 0.3744 & 59.1 & 4.74 & 3.71 \\
PBEsol     & 3.996 & 6.590 & 1.6492 & 0.3744 & 67.6 & 4.81 & 4.14 \\
AM05       & 3.999 & 6.594 & 1.6490 & 0.3745 & 66.0 & 4.95 & 3.91 \\
PBEsol+$U$ & 3.980 & 6.568 & 1.6503 & 0.3743 & 69.0 & 4.80 & 4.13 \\
AM05+$U$   & 3.980 & 6.568 & 1.6503 & 0.3743 & 67.4 & 4.81 & 3.89 \\ \hline

Exp.       & 3.96\footnote{Potentially strained samples from micro-indentation~\cite{Xiao.Pirouz:1992:JoMR}.} 
           & 6.57\footnotemark[1] 
           & 1.659\footnotemark[1] \\
           & 3.9878(20)\footnote{Room-temperature x-ray diffraction of unstrained crystalline nanowires \cite{Ziss.Stangl:2018}.}
           & 6.5776(3)\footnotemark[2] 
           & 1.6494\footnotemark[2] \\
\end{tabular}
\end{ruledtabular}
\end{table*}

The positions of the four atoms in the unit cell of the lonsdaleite structure are defined by the hexagonal lattice constants $a$ and $c$, as well as the dimensionless internal cell parameter $u$. In Table~\ref{tab:hex_ground}, the results of our calculations of the structural properties of 2H-Ge with various XC functionals are compiled. As already discussed for 3C-Ge, we observe a consistent over- and underestimation of bond lengths depending on the choice of the functional. These tendencies are also reflected in the values for the cohesive energy $E_\mathrm{coh}$ and the bulk modulus $B_0$. 

On average, the bond lengths of 2H-Ge are slightly reduced in comparison to the 3C-Ge bond length $d=\sqrt{3}/4\,a_0$. This can be understood in terms of a detailed analysis of the atomic geometry. The Ge-Ge bonds parallel to the $c$~axis (bond length $d_\|=u\,c$) and those not parallel to the $c$~axis (bond length $d_\perp=\sqrt{a^2/3+(1/2-u)^2c^2}$) form distorted tetrahedra. All bonds in the distorted tetrahedra have the same length only when $u=1/4+1/3\,(c/a)^{-2}$ holds. For the ideal lonsdaleite structure with $(c/a)_\mathrm{ideal}=\sqrt{8/3}$ and $u_\mathrm{ideal}=3/8$, also the bond angles become equal and the coordination tetrahedra are regular. 

We can deduce from Table~\ref{tab:hex_ground} that $c/a>(c/a)_{\rm ideal}$ and $u<u_{\rm ideal}$. The relation $d_\perp<d<d_\|$ holds, resulting in tetrahedra that are slightly elongated along the $c$~axis. For instance, we find $d_\perp=2.451~\Angst$ and $d_\|=2.468~\Angst$ for the PBEsol functional. The average bond length $d_\mathrm{av}=2.455~\Angst$ of 2H-Ge is only slightly smaller than the 3C-Ge bond length $d=2.457~\Angst$. These findings are in line with the empirical rule of Lawaetz~\cite{Lawaetz:1972:PRB} for III-V compounds, which states that materials with $c/a>(c/a)_{\rm ideal}$ favor a zincblende ground-state structure, or for an elemental materials as Ge, the diamond structure. The computed values of $u$ nearly follow the relation $u=1/4+1/3\,(c/a)^{-2}$ indicating that the deformation of bonding tetrahedra in 2H-Ge can be explained to a good share by deviations of the bond angles from the ideal value. 

In summary, we observe a relatively strong hexagonal crystal deformation in 2H-Ge which is characterized by large $(c/a)-(c/a)_\mathrm{ideal}$ and $u-u_\mathrm{ideal}$, despite the presence of covalent bonds. The calculated lattice parameters of 2H-Ge agree very well with the available experimental data \cite{Xiao.Pirouz:1992:JoMR,Ziss.Stangl:2018}. However, experimental structural parameters are scarce in the literature and, in some cases, have been obtained from nano-structured and potentially strained samples.


\subsection{Electronic structure}

\begin{figure*}
\includegraphics[width=0.49\linewidth]{{fig3a}}
\hfill
\includegraphics[width=0.49\linewidth]{{fig3b}}
\caption{\label{fig:hex_bst} Band structure of 2H-Ge computed with the HSE06 and MBJLDA functionals (a). The irreducible representations of relevant high-symmetry states in the band-gap region are given in the double-group notation of Koster \emph{et al.}~\cite{Koster.Dimmock.ea:1963:Book} (b). The VBM is set to zero.}
\end{figure*}

\begin{table}[t] 
 \caption{\label{tab:hex_eigenvalues} Band energies of 2H-Ge computed with different XC functionals. The $\Gamma_{9v}^+$ VBM is used as energy zero. The crystal-field and spin-orbit splitting parameters $\Delta_\cf$, $\Delta_\so^\|$, and $\Delta_\so^\perp$ for the VBM have been calculated from the band energies. Band energies from an empirical-pseudopotential model (EPM) \cite{De.Pryor:2014:JoPCM} are given for comparison. All energies in eV.}
\begin{ruledtabular}
\begin{tabular}{lccc}
State  & HSE06 & MBJLDA & EPM (Ref.~\cite{De.Pryor:2014:JoPCM}) \\ \hline
$\Gamma_{7v-}^+$ & -0.484 & -0.433 & -0.490 \\
$\Gamma_{7v+}^+$ & -0.134 & -0.120 & -0.129 \\
$\Gamma_{9v}^+$  &  0.000 &  0.000 &  0.000 \\
$\Gamma_{8c}^-$  &  0.286 & 0.298  &  0.310 \\
$\Gamma_{7c}^-$  &  0.614 & 0.632  &  0.766 \\
$U_{5c}$         &  0.615 & 0.620  &        \\ \hline
$\Delta_\cf$     &  0.288 & 0.270 & \\
$\Delta_\so^\|$  &  0.329 & 0.282 & \\
$\Delta_\so^\perp$& 0.320 & 0.274 & \\
\end{tabular}
\end{ruledtabular}
\end{table}

The band structure of 2H-Ge including SOC has been calculated with the HSE06 and MBJLDA functionals for the PBEsol atomic structure (see Fig.~\ref{fig:hex_bst}). As for 3C-Ge, local (LDA) and semilocal (PBE, PBEsol, AM05) XC functionals yield negative band gaps. High-symmetry states in the vicinity of the band gap are labeled according to the double-group notation of Koster \emph{et al.}~\cite{Koster.Dimmock.ea:1963:Book}. Their energies are also given in Table~\ref{tab:hex_eigenvalues}. The state labels are essentially the same as for the wurtzite structure. Due to the additional inversion symmetry in the lonsdaleite crystal structure, some high-symmetry states also have a well-defined parity which is indicated by a superscript index. We explicitly calculated the parities of the high-symmetry states at  time-reversal invariant momentum (TRIM) points (see Fig.~\ref{fig:hex_bst} and Table~\ref{tab:hex_eigenvalues}) that are characterized by the relation $\kv_\mathrm{TRIM} = -\kv_\mathrm{TRIM}+\Gv$ ~\cite{Fu.Kane:2007:PRB}. At the TRIM points, a displacement by a reciprocal lattice vector $\Gv$ reverses the application of time-reversal symmetry. In the lonsdaleite structure, these TRIM points are the $\Gamma$, the 3 $L$, and the 3 $M$ points of the hexagonal BZ. The parities we find are partially at odds with those given by De \emph{et al.}~\cite{De.Pryor:2014:JoPCM}. However, the optical oscillator strengths that we obtain for the near-gap transitions (see below) corroborate our findings. Moreover, our parities are in line with the results of Salehpour \emph{et al.}~\cite{Salehpour.Satpathy:1990:PRB} for carbon in the lonsdaleite structure.

We find 2H-Ge to be a direct-gap semiconductor with a band gap of 0.286~eV (HSE06) or 0.298~eV (MBJLDA). Note that the precise gap values are outmost sensitive to lattice strain due to the huge deformation potentials of the gap-forming states. Our results agree well with previously calculated gaps of 0.31~eV (empirical-pseudopotential method~\cite{De.Pryor:2014:JoPCM}) or 0.32~eV (HSE06 calculation~\cite{Kaewmaraya.Vincent.ea:2017:JPCC}), being slightly higher than a $GW$ gap of 0.23~eV obtained by Chen \emph{et al.}~\cite{Chen.Fan.ea:2017:JoPDAP}. As for 3C-Ge, the HSE06 and MBJLDA energies of the near-gap states match excellently. Deviations occur for states further away from the band-gap region (see Fig.~\ref{fig:hex_bst}).

The conduction-band minima of 3C-Ge are the four $L_{6c}^+$ states. The $L_{6c}^+$ state in $[111]$ direction is backfolded to the $\Gamma_{8c}^-$ state in 2H-Ge and becomes the CBM of 2H-Ge. The other $L$~points of 3C-Ge are mapped onto a point between $M$ and $L$ on the $U$ line of the hexagonal BZ. In the ideal lonsdaleite structure, the backfolded $L$ points lie at $\frac{2}{3}\overline{ML}$, whereas they are slightly shifted along the $U$ line in the relaxed structure~\cite{Cohen.Chelikowsky:1988:Book}. Indeed, we observe a minimum of the first conduction band on the $U$ line of 2H-Ge at approximately $\frac{2}{3}\overline{ML}$ that is almost degenerate in energy with the $L_{6c}^+$ state of 3C-Ge.

The $\Gamma_{8c}^-$ state in 2H-Ge is downshifted by 0.4~eV compared to the cubic $L_{6c}^+$ state which cannot be understood by simple folding arguments. It is, however, in agreement with the behavior of Si going from diamond to lonsdaleite structure~\cite{Raffy.Furthmueller.ea:2002:PRB,Roedl.Sander.ea:2015:PRB}, whereas it is in clear contrast with the small band-gap opening in biatomic semiconductors when the structure changes from zincblende to wurtzite~\cite{Bechstedt.Belabbes:2013:JoPCM}. The cubic $\Gamma_{7c}^-$ state coincides with the $\Gamma_{7c}^-$ state of the second conduction band in the lonsdaleite structure. 
 
Due to the presence of inversion symmetry in the lonsdaleite structure, the valence bands in 2H-Ge do not show a spin-orbit-induced splitting of the $\kv$~dispersion along the $\Gamma$-$M$ line as it occurs in wurtzite semiconductors~\cite{Bechstedt.Belabbes:2013:JoPCM,Carvalho.Schleife.ea:2011:PRB}. Following $\kv\cdot\pv$ theory, we can write the energy splittings at $\Gamma$~\cite{Chuang.Chang:1996:PRB} as
\begin{equation}\label{eq:splittings}
\begin{split}
\eps(\Gamma_{9v}^+)-\epsilon(\Gamma_{7v+/-}^+)& =\frac{\Delta_\cf+\Delta_\so^\|}{2} \\
&\mp \frac{1}{2}\sqrt{\left(\Delta_\cf-\frac{1}{3}\Delta_\so^\|\right)^2+\frac{8}{9}(\Delta_\so^\perp)^2}.
\end{split}
\end{equation}
These formulas allow to extract the crystal-field splitting $\Delta_\cf$ and the spin-orbit splitting parameters parallel and perpendicular to the $c$~axis, $\Delta_\so^\|$ and $\Delta_\so^\perp$. The band ordering $\Gamma_{9v}^+>\Gamma_{7v+}^+>\Gamma_{7v-}^+$ at the top of the valence bands which we find for 2H-Ge is in line with the ordering observed in wurtzite semiconductors (except AlN and ZnO for which the band ordering is $\Gamma_{7v+}>\Gamma_{9v}>\Gamma_{7v-}$)~\cite{Carvalho.Schleife.ea:2011:PRB,Schleife.Rodl.ea:2007}. Note that the subscript indices $7v\pm$ represent bands of the same symmetry; the symbols $\pm$ merely serve to distinguish between the upper and the lower state. They are not to be confused with superscript parity indices. 

We have extracted the crystal-field splitting from a calculation without SOC and used Eq.~\eqref{eq:splittings} to compute the spin-orbit splitting parameters from the band splittings of the calculation including SOC. The resulting values are compiled in Table~\ref{tab:hex_eigenvalues}. In particular, the direction-averaged spin-orbit splitting $\Delta_\so=(\Delta_\so^\|+2\Delta_\so^\perp)/3$ compares well to the spin-orbit splitting of the VBM in 3C-Ge (cf.\ Table~\ref{tab:cub_eigenvalues}). The crystal-field splitting in 2H-Ge is much larger than for  III-V compounds that crystallize in zincblende or wurtzite structure under ambient conditions~\cite{Bechstedt.Belabbes:2013:JoPCM}. The large crystal-field splitting for 2H-Ge is, however, in accordance with the significant deformation of the bonding tetrahedra, as indicated by the increase of $c/a$ (see Table~\ref{tab:hex_ground}) with respect to its ideal value. The large $\Delta_\cf$ shifts the $\Gamma_{9v}^+$ level toward higher energies and, hence, explains the observed small direct gap. We emphasize that the quasicubic approximation $\Delta_\so^\|=\Delta_\so^\perp$ that was used by De \emph{et al.}~\cite{De.Pryor:2014:JoPCM} is not valid for 2H-Ge and leads to splitting parameters at variance with our values.

\begin{table}[t]
\caption{\label{tab:hex_effmass} Effective electron and hole masses of 2H-Ge in  units of the free electron mass $m$. The masses are given for several directions in the BZ.  The VBM at $\Gamma$ splits into a heavy hole ($m_\mathrm{h}^\mathrm{hh}$), light hole ($m_\mathrm{h}^\mathrm{lh}$), and split-off hole ($m_\mathrm{h}^\so$).}
\begin{ruledtabular}
\begin{tabular}{llcc}
Mass & Direction & HSE06 & MBJLDA \\ \hline
$m_\mathrm{h}^\so(\Gamma_{7v-}^+)$ 
& $\Gamma\to M$ & 0.252 & 0.325 \\
& $\Gamma\to A$ & 0.044 & 0.053 \\
$m_\mathrm{h}^\mathrm{lh}(\Gamma_{7v+}^+)$ 
& $\Gamma\to M$ & 0.079 & 0.101 \\
& $\Gamma\to A$ & 0.085 & 0.120 \\
$m_\mathrm{h}^\mathrm{hh}(\Gamma_{9v}^+)$ 
& $\Gamma\to M$ & 0.055 & 0.074 \\
& $\Gamma\to A$ & 0.463 & 0.526 \\  \hline
$m_\mathrm{e}(\Gamma_{8c}^-)$ 
& $\Gamma\to M$ & 0.076 & 0.089 \\
& $\Gamma\to A$ & 0.997 & 1.088 \\
$m_\mathrm{e}(\Gamma_{7c}^-)$ 
& $\Gamma\to M$ & 0.038 & 0.052 \\
& $\Gamma\to A$ & 0.033 & 0.042 \\
\end{tabular}
\end{ruledtabular}
\end{table}

The effective masses of the band edges at $\Gamma$ are compiled in Table~\ref{tab:hex_effmass}. The small electron mass with almost vanishing anisotropy for the $\Gamma_{7c}^-$ conduction band of 2H-Ge is of the order of magnitude of the $\Gamma_{7c}^-$ mass of 3C-Ge in Table~\ref{tab:cub_effmass}. The mass tensor at the CBM $\Gamma_{8c}^-$, on the other hand, is highly anisotropic with a large mass along the hexagonal $c$~axis and a small mass in the plane perpendicular to it. These values qualitatively agree with the longitudinal and transverse masses $m_\mathrm{e}^\|(L_{6c}^+)$ and $m_\mathrm{e}^\perp(L_{6c}^+)$ at the $L_{6c}^+$ minimum of 3C-Ge. The strong direction dependence of the $\Gamma_{8c}^-$ conduction-band dispersion is consistent with the identification of the band symmetry. The hole masses also exhibit strong asymmetries, especially for the $\Gamma_{9v}^+$ and $\Gamma_{7v-}^+$ bands. 


\subsection{Optical transitions}
\label{subsec:optics}

\begin{table*}[t]
\caption{\label{tab:hex_optmat} Optical transitions between valence and conduction bands of 2H-Ge characterized by transition energy, optical transition matrix element, Kane energy, and oscillator strength. Transitions that are dipole forbidden by symmetry are indicated by horizontal lines. The values have been calculated with the HSE06 and MBJLDA functionals for light polarized perpendicular and parallel to the $c$ axis.}
\begin{ruledtabular}
\begin{tabular}{llccccccc}
Transition & Method & Transition energy & \multicolumn{2}{c}{Optical transition matrix element} & \multicolumn{2}{c}{Kane energy} & \multicolumn{2}{c}{Oscillator strength} \\
& & $\eps_{c\kv}-\eps_{v\kv}$ (eV) & $p^\perp$ ($\hbar/\aB$) & $p^\|$ ($\hbar/\aB$) & $E_p^\perp$ (eV) & $E_p^\|$ (eV) & $f^\perp$ & \multicolumn{1}{c}{$f^\|$} \\ \hline
$\Gamma_{9v}^+ \to \Gamma_{8c}^-$ 
& HSE06  & 0.286 & $6.48\cdot10^{-3}$ & --- & $2.29\cdot10^{-3}$ & --- & $8.00\cdot10^{-3}$ & --- \\
& MBJLDA & 0.298 & $5.92\cdot10^{-3}$ & --- & $1.91\cdot10^{-3}$ & --- & $6.39\cdot10^{-3}$ & --- \\
$\Gamma_{7v+}^+ \to \Gamma_{8c}^-$ 
& HSE06  & 0.419 & --- & --- & --- & --- & --- & --- \\
& MBJLDA & 0.418 & --- & --- & --- & --- & --- & --- \\
$\Gamma_{7v-}^+ \to \Gamma_{8c}^-$ 
& HSE06  & 0.770 & --- & --- & --- & --- & --- & --- \\
& MBJLDA & 0.730 & --- & --- & --- & --- & --- & --- \\ \hline
$\Gamma_{9v}^+ \to \Gamma_{7c}^-$  
& HSE06  & 0.614 & 0.447 & --- & 10.9 & --- & 17.7 & --- \\
& MBJLDA & 0.632 & 0.394 & --- & 8.44 & --- & 13.4 & --- \\
$\Gamma_{7v+}^+ \to \Gamma_{7c}^-$ 
& HSE06  & 0.748 & 0.384 & 0.388 & 8.01 & 8.21 & 10.7 & 11.0 \\
& MBJLDA & 0.752 & 0.343 & 0.330 & 6.42 & 5.92 & 8.54 & 7.87 \\
$\Gamma_{7v-}^+ \to \Gamma_{7c}^-$ 
& HSE06  & 1.098 & 0.214 & 0.661 & 2.49 & 23.8 & 2.27 & 21.7 \\
& MBJLDA & 1.065 & 0.178 & 0.603 & 1.72 & 19.8 & 1.62 & 18.6 \\
\end{tabular}
\end{ruledtabular}
\end{table*}

\begin{figure}
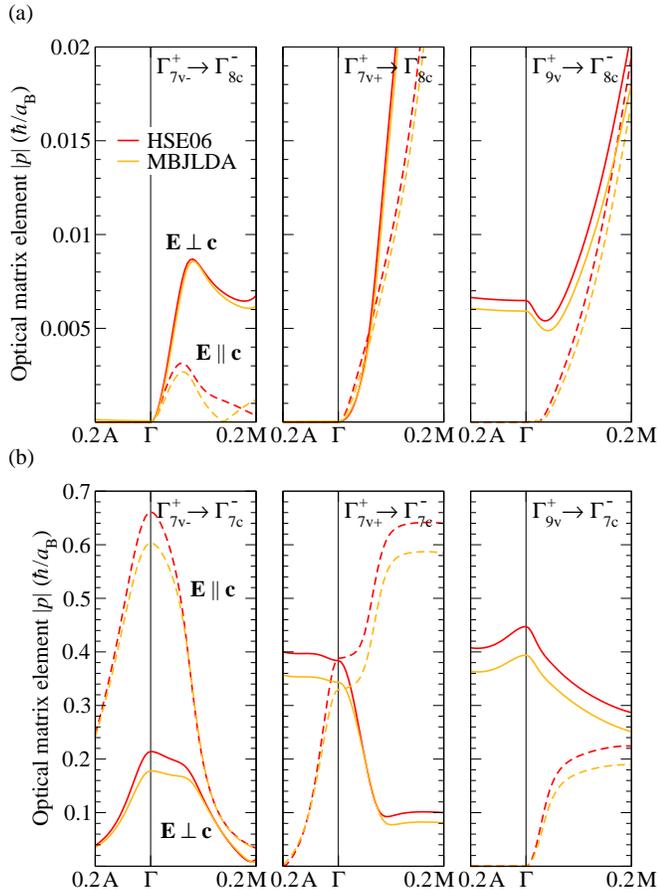

\includegraphics[width=\linewidth]{{fig4a}}
\includegraphics[width=\linewidth]{{fig4b}}
\caption{\label{fig:hex_optmat} Optical transition matrix elements $|p|$ of the lowest band-to-band transitions in 2H-Ge along high-symmetry lines in the vicinity of $\Gamma$. The matrix elements have been calculated for ordinary ($\Ev\perp\mathbf{c}$) and extraordinary ($\Ev||\mathbf{c}$) light polarization using the HSE06 and the MBJLDA functionals, respectively.}
\end{figure}

\begin{figure}
\includegraphics[width=\linewidth]{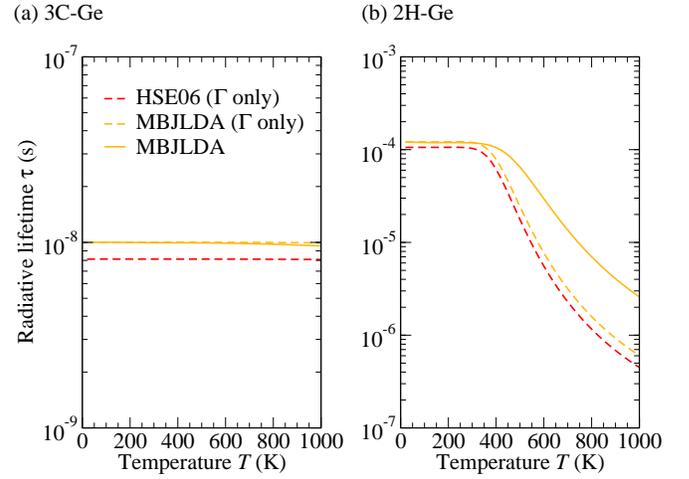}
\caption{\label{fig:hex_tau} Radiative lifetime $\tau$ versus temperature for 3C-Ge and 2H-Ge calculated at the $\Gamma$ point only (dashed lines) or integrating over the whole BZ (solid lines). Results are given for the HSE06 and MBJLDA functionals.}
\end{figure}

The oscillator strengths of optical transitions between the three uppermost valence and two lowest conduction bands of 2H-Ge are given in Table~\ref{tab:hex_optmat}. Transitions that are dipole forbidden due to group-theoretical arguments~\cite{Tronc.Kitaev.ea:1999:pssb} are indicated by horizontal lines. These symmetry considerations corroborate our identified band ordering at the $\Gamma$ point ($\Gamma_{7c}^->\Gamma_{8c}^-)$. In Fig.~\ref{fig:hex_optmat}, the corresponding optical matrix elements are plotted along high-symmetry lines close to $\Gamma$. It is evident that transitions which are dipole forbidden at $\Gamma$ can be dipole allowed in its immediate vicinity. As a direct consequence, a transition that is dipole forbidden at $\Gamma$ at zero temperature may become optically active at higher temperatures when electrons and holes populate the bands also in the surroundings of $\Gamma$. What is more, deviations from the perfect lonsdaleite structure due to defects, nanostructuring, or surfaces/interfaces may violate the $\kv$-selection rule rendering beforehand dipole-forbidden transitions dipole allowed.

Lonsdaleite Ge, being a direct semiconductor with a very weak lowest optical transition, exhibits significant variations in luminescence and absorption in comparison to cubic Ge. The effects can be expected to be stronger than for Si~\cite{Roedl.Sander.ea:2015:PRB} and SiGe alloys~\cite{Amato.Kaewmaraya.ea:2016:NL}. For a clear illustration of the global light-emission properties, the radiative lifetime $\tau$ as a function of temperature is shown in Fig.~\ref{fig:hex_tau}. The lifetimes were calculated according to Eq.~\eqref{eq:av_rec_rate} using transition energies and optical matrix elements obtained with the HSE06 and the MBJLDA functionals which yield comparable results. Full $\kv$-point convergence could, however, only be achieved with the computationally cheaper MBJLDA functional.

When we compare the radiative lifetime of 3C-Ge and 2H-Ge, we find striking differences. As expected, the radiative lifetime of cubic Ge is largely temperature independent because of its larger fundamental band gap. Excluding off-$\Gamma$ optical transitions in the evaluation of the lifetime does not have any significant impact. The radiative lifetime of lonsdaleite Ge, on the other hand, which is very high at low temperatures due to the extremely weak oscillator strength of the lowest $\Gamma$-$\Gamma$ transition, decreases rapidly above 400~K, when the second conduction band that is optically active starts to be populated. What is more, off-$\Gamma$ optical transitions significantly contribute to the lifetime. This is easily understood by recalling that optical matrix elements that vanish by symmetry at $\Gamma$ can be non-zero in the immediate vicinity of the BZ center (see Fig.~\ref{fig:hex_optmat}). 

The large gap difference between 3C-Ge and 2H-Ge and its consequence for the temperature-dependent band populations explain the huge difference of $\tau$ by several orders of magnitude for low temperatures. (Note that for the thermalization of electrons and holes in Ge nanocrystals similar curves have been published~\cite{Weissker.Ning.ea:2011:PRB}.) Manipulation of the atomic structure of 2H-Ge by straining or alloying, for instance, may lead to an inversion of the $\Gamma_{8c}^-$ and $\Gamma_{7c}^-$ conduction states which is likely to drastically improve the light-emission properties of lonsdaleite Ge, thus providing a vast playground for engineering its optoelectronic performance.


\section{Summary and Conclusions}
\label{sec:summary}

The lonsdaleite (2H) phase of Ge, which can be grown using hexagonal III-V nanowire templates, is considered a good candidate for Si on-chip optical interconnects and Si-compatible quantum light sources, thanks to its predicted direct band gap. Since experimental data and reliable calculations on 2H-Ge are scarce and often inconsistent, we first established our computational approach for efficient predictive {\it ab initio} calculations in this work. We systematically benchmarked the performance of several XC functionals of DFT, including meta-GGA and hybrid functionals, to calculate the experimentally and theoretically well known structural and electronic properties of diamond-structure (3C) Ge. In a second step, we used these functionals to predict the structural, electronic, and optical properties of lonsdaleite Ge.

The atomic structure of 2H-Ge was computed with the PBEsol functional which is shown to yield excellent lattice parameters for the well studied cubic phase of Ge. The electronic structures of cubic and lonsdaleite Ge were calculated with the HSE06 hybrid functional and the MBJLDA meta-GGA, finding consistent results with both approaches, and an excellent agreement with the available experimental data. The $\Gamma_{8c}^-$ CBM of lonsdaleite Ge results from the backfolding of the $L$ point of diamond-structure Ge onto the $\Gamma$ point of the hexagonal BZ, while the $\Gamma_{7c}^-$ conduction-band state, that is derived from the lowest conduction band at $\Gamma$ of cubic Ge, is pushed towards higher energies. The energetic ordering of the three highest valence bands is $\Gamma_{9v}^+>\Gamma_{7v+}^+>\Gamma_{7v-}^+$. While the spin-orbit splittings of the hexagonal and cubic phase are similar, a huge crystal-field splitting is observed in 2H-Ge. The crystal-field splitting is responsible for the small $\Gamma_{9v}^+\rightarrow\Gamma_{8c}^-$ band gap of only about 0.3~eV. The second CBM $\Gamma_{7c}^-$ is higher in energy by about 0.3~eV. The calculated electron and hole effective masses of cubic Ge are in good agreement with values in literature. Consequently, we expect to predict reliable effective masses for electrons and holes in 2H-Ge. 

The dipole-allowed and dipole-forbidden optical transitions between the uppermost valence bands and lowest conduction bands near the $\Gamma$ point and their polarization dependence is consistent with the symmetry identification of the bands. We prove that lonsdaleite Ge is a semiconductor with a direct fundamental gap in the infrared which exhibits a non-vanishing but small optical oscillator strength only for ordinary light polarization. The optical transitions to the second lowest conduction band instead are dipole allowed with large oscillator strengths. We notice that the distance between the first and second conduction band, as well as the size of the band gap, appear to be sensitive to the structural parameters. Consequently, a careful investigation of the luminescence properties, including their time dependence, and the absorption edge, also considering effects of strain, are suggested to further clarify the optical and optoelectronic properties of the promising new material lonsdaleite Ge.

\begin{acknowledgments}
We thank the FET Open project SiLAS (GA No.\ 735008) of the European Commission for financial support. We are grateful for fruitful discussions with the members of the SiLAS collaboration. C.\,R. acknowledges financial support from the Marie Sk\l{}odowska-Curie Actions (GA No.\ 751823). Computational resources were provided by the Leibniz Supercomputing Centre on SuperMUC (project No.\ pr62ja). 
\end{acknowledgments}


%

\end{document}